\documentclass[twocolumn,showpacs,preprintnumbers,amsmath,amssymb,aps,superscriptaddress]{revtex4-1}

\usepackage{graphicx}
\usepackage{eurosym}
\usepackage[usenames,dvipsnames,table]{xcolor}
\usepackage[utf8]{inputenc}
\usepackage{lmodern}
\usepackage{physics}
\usepackage{amsmath}
\usepackage{bm}
\usepackage{feynman}
\usepackage{slashed}
\usepackage{comment}
\usepackage[T1]{fontenc} 
\usepackage{mathtools}
\usepackage[breaklinks,colorlinks,urlcolor=blue,citecolor=blue,linkcolor=blue]{hyperref}

\newcommand{\m}[1]{\mathcal{#1}}

\begin{document}

\title{Color Coherence Effects in Dipole-Quark Scattering in the Soft Limit}

\author{Daniel Pablos}
\email{daniel.pablos@usc.es}
\affiliation{Instituto Galego de F\'isica de Altas Enerx\'ias IGFAE, Universidade de Santiago de Compostela, E-15782 Galicia-Spain}
\author{Sergio Sanjurjo}
\email{sergiosanjurjom@gmail.com}
\affiliation{Departamento de F\'isica, Universidad de Oviedo, Avda. Federico Garc\'ia Lorca 18, 33007 Oviedo, Spain}
%

\date{\today}

\begin{abstract}
Color coherence effects play a crucial role in the description of jet evolution at collider experiments. It is well known that the stimulated gluon emission suffered by energetic jets traversing deconfined QCD matter is also affected by color coherence effects. Through multiple soft scatterings with the medium constituents, an antenna will lose its color correlation, causing its legs to behave as independent emitters after the so-called decoherence time. In this work we provide the first computation of the properties of the recoils produced as a result of these soft scatterings between a color coherent dipole and the medium constituents. Our findings reveal that the angular phase-space of these soft recoils is strongly restricted by the opening angle of the dipole itself due to quantum interference effects. In this long wavelength limit, one can effectively consider that interactions take place with each of the legs of the dipole separately, provided that the angular constraints dictated by the corresponding color flow topologies are respected. This is in complete analogy with the case of soft gluon emission in vacuum, where the recoil quark plays the role of the emitted gluon. As a direct phenomenological application we estimate the collisional energy loss rate of a color antenna. Importantly, these results indicate the way in which color coherence effects can be implemented in jet quenching models that account for the recoils from elastic scatterings, improving in this way our description of medium response physics in heavy-ion collisions.
\end{abstract}

\maketitle

\section{Introduction}
Color coherence phenomena are present in any gauge theory. In Quantum Chromodynamics (QCD), experimental confirmation of the detailed predictions associated to color flow topologies has contributed to validating the perturbative approach to jet physics at colliders. Salient examples include the phenomenon of drag, or string, effects~\cite{Dokshitzer:1987nm,Dokshitzer:1991wu} in which the multiplicities of two- and three-jet events depend on the geometry of those jet configurations, and also the phenomenon of angular ordering (AO) in soft gluon emission~\cite{Mueller:1981ex,Ermolaev:1981cm}, leading to a depletion of soft particles yields around a jet due to destructive interferences, known as the humpback plateau~\cite{Azimov:1985by}.
The remarkable AO effects state that soft gluon emission off a color dipole can be effectively described as the independent emission off each of the legs of the dipole provided that one observes certain restrictions in the angular phase-space of the emission. This picture provides the basis of the probabilistic approach to the simulation of jet evolution used in modern event generators~\cite{Marchesini:1983bm}.

The evolution of jets traversing deconfined QCD matter, such as the quark-gluon plasma (QGP) created in heavy-ion collisions~\cite{Busza:2018rrf}, has also been shown to be affected by coherence effects. Stimulated gluon emission via repeated scatterings with the medium constituents~\cite{Baier:1996kr,Baier:1996sk,Zakharov:1996fv,Zakharov:1997uu,Gyulassy:2000fs,Gyulassy:2000er,Wiedemann:2000za,Wang:2001ifa,Djordjevic:2003zk} leads to a degradation of the jet energy and a modification of its substructure, a set of phenomena commonly referred to as jet quenching~\cite{dEnterria:2009xfs,Wiedemann:2009sh,Majumder:2010qh,Mehtar-Tani:2013pia}. The study of the bremsstrahlung pattern off a color dipole~\cite{Mehtar-Tani:2010ebp,Casalderrey-Solana:2011ule,Mehtar-Tani:2011hma,Mehtar-Tani:2011lic,Mehtar-Tani:2012mfa,Casalderrey-Solana:2015bww} with opening angle $\theta$ revealed that each leg will behave as an independent emitter only after the decoherence time $\tau_d\sim \theta^{-2/3}$, when independent color rotations have led to a loss of the color coherence of the pair. 
While much is known about the description of medium-induced radiation off a color dipole and its phenomenological consequences~\cite{Casalderrey-Solana:2012evi,Mehtar-Tani:2016aco,Mehtar-Tani:2017web,Hulcher:2017cpt,Caucal:2018dla,Caucal:2019uvr,Casalderrey-Solana:2019ubu,Caucal:2020xad,Mehtar-Tani:2021fud,Caucal:2021cfb,Pablos:2022mrx,Belmont:2023fau,Andres:2022ovj,Andres:2023xwr,Cunqueiro:2023vxl,Barata:2023bhh,Mehtar-Tani:2024jtd}, this is not the case for the description of the medium constituents after the multiple soft scatterings take place. This is largely due to the fact that the importance of these recoils has only been acknowledged relatively recently (see~\cite{Cao:2020wlm} and references therein). Interesting on their own right as a means to learn about the inner workings of the QGP at various length scales~\cite{DEramo:2012uzl,DEramo:2018eoy,Kumar:2019uvu,Yang:2023dwc}, they appear to be crucial in the description of most jet observables~\cite{Park:2018acg,JETSCAPE:2018vyw,Milhano:2022kzx,JETSCAPE:2022jer,JETSCAPE:2023hqn,Luo:2023nsi}.
They contribute as correlated background, influencing the way we interpret jet quenching phenomena in experiments and thus affecting our ability to understand the various physical mechanisms at play.

This Letter addresses, for the first time, the coherence effects on the properties of the medium recoils involved in scattering with a color-coherent energetic dipole. 
Crucially, highly energetic partons produced during jet evolution \emph{cannot} be considered in general as on-shell states coming from infinity, and their correct description is in terms of a collection of color correlated dipoles.
A number of jet quenching models already account for the presence of the recoils from elastic scatterings~\cite{He:2015pra,Zapp:2013vla,Ke:2020clc,JETSCAPE:2021ehl,Hulcher:2022kmn}, but do not account for color coherence effects whatsoever as they work using on-shell states. 

The rest of the manuscript is organized as follows. Using standard perturbative techniques, we show in Section~\ref{sec:matrix} that the angular distribution of the soft recoils is strongly influenced by interference phenomena, closely resembling the patterns found for soft gluon emission. These results allow us to compute the first estimate for the collisional energy loss rate of a color dipole in Section~\ref{sec:collloss}. Finally, in Section~\ref{sec:conc} we look ahead and comment on the implications of these results in the modelling of elastic collisions in jet quenching event generators.

\section{Color dipole scattering with a quark in the soft limit}
\label{sec:matrix}

\begin{figure}[t!]
    \includegraphics[width=0.235\textwidth]{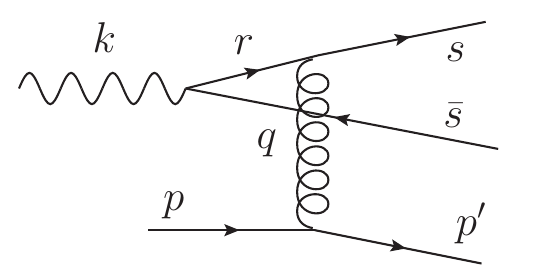}
    \includegraphics[width=0.235\textwidth]{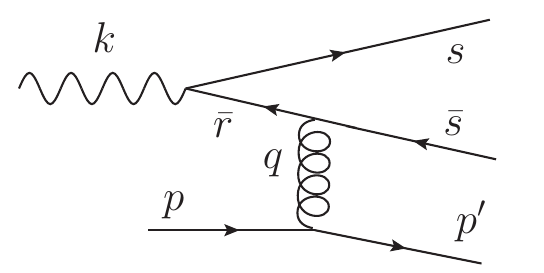}
    \caption{The only two leading diagrams ($i\m{M}_1^{\gamma}$ in the left, $i\m{M}_2^{\gamma}$ in the right) for the scattering between a color singlet dipole and a quark in the soft limit.}
    \label{fig:diag}
\end{figure}

We analyze the scattering process between a soft quark and a virtual photon that decays into a $q\bar{q}$ pair.
The quark is soft both regarding its rest mass $m$ and the energy $E$ acquired in the scattering (assuming $m\ll E$) with the energetic dipole with total energy $w$. 
We consider those diagrams that are leading in $\beta=E/w$. 
The two leading diagrams are those with a scattering with either the quark or the anti-quark by the exchange of a virtual gluon in the $t$-channel. All other diagrams ($s$-channel as well as $u$-channel contributions) are suppressed in the soft limit we are interested in, i.e. they are subleading in $\beta$. Using Feynman rules in momentum space, the $t$-channel diagrams read (see Fig.~\ref{fig:diag})
\begin{align}
i \m{M}_1^{\gamma}&=\frac{-i g_s^2 e Q}{q^2} a_{ij}a_{kl} \, \left[\bar{u}(s)\gamma^{\mu}\frac{\slashed r}{r^2} \Gamma v(\bar{s})\right]
\left[\bar{u}(p')\gamma_{\mu}u(p)\right] \\
i \m{M}_2^{\gamma}&=\frac{-i g_s^2 e Q}{q^2} a_{ij}a_{kl} \, \left[\bar{u}(s)\Gamma\frac{-\slashed{\bar{r}}}{\bar{r}^2} \gamma^{\mu}v(\bar{s})\right]
\left[\bar{u}(p')\gamma_{\mu}u(p)\right] \, .
\end{align}
The term $\Gamma$ represents the interaction vertex associated to the origination of the dipole.
$eQ$ is the electric charge carried by the quarks of the antenna and $g_s$ is the strong coupling constant. The matrix $a$ is the generator of the $SU(3)$ color group and the latin indices $ij$ and $kl$ in the fundamental representation indicate the color flow in the soft quark and antenna, respectively. We are working in the Feynman gauge.

Regarding the algebra involving the energetic legs of the dipole, $s$ and $\bar{s}$, we assume that the exchanged momentum is small and use the equations of motion for a massless quark, so $\slashed r u(s) \approx \slashed s u(s)=0$, and analogously for the anti-quark. The sum of the two matrix elements is (using $\lbrace \gamma^{\mu},\gamma^{\nu}\rbrace=2 g^{\mu \nu}$)
\begin{align}
\label{eq:photmatrix}
&i \m{M}^{\gamma}  =  i \m{M}_1^{\gamma}+i \m{M}_2^{\gamma}= \frac{-2 g_s^2}{q^2}a_{ij}a_{kl}B^{\mu}\left( \frac{s_{\mu}}{r^2}-\frac{\bar{s}_{\mu}}{\bar{r}^2}\right)  \, \Gamma^{\gamma}_{q\bar{q}}\, ,
\end{align}
where we have defined $B^{\mu}\equiv\bar{u}(p')\gamma^{\mu}u(p)$. Here, $\Gamma^{\gamma}_{q\bar{q}}\equiv i e Q \,  \bar{u}(s)\Gamma v(\bar{s})$, and its explicit form is not relevant for our present discussion as it is factorized from the rest of the expression in the soft limit.
Averaging over initial color states ($N$ possible colors for the incoming soft quark) and summing over final color states, the color factor gives
\begin{align}
&\frac{1}{N}\sum (a_{ij}a_{kl})^* \, (b_{ij}b_{kl})=\frac{C_F}{2} \, .
\end{align}

\begin{figure}[t!]
    \includegraphics[width=0.35\textwidth]{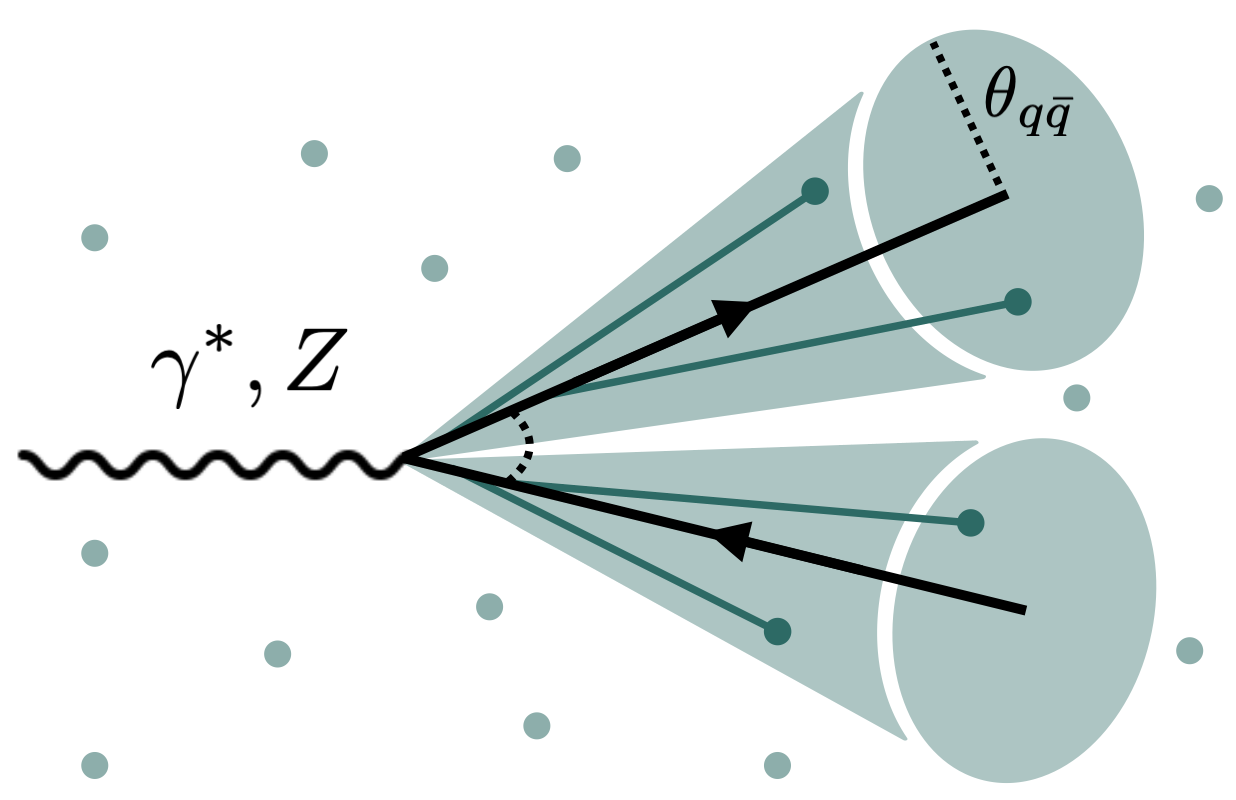}
    \caption{Sketch representing a color singlet dipole (solid black lines) produced by the decay of a colorless boson (wiggly black line) that interacts with some medium constituents (green dots). Quantum interference effects restrict the recoil angles (green lines ending with dots) to lie within cones of the angular size of the dipole, $\theta_{q \bar{q}}$ (dotted black lines), centered at each of the dipole legs.}
    \label{fig:sketch}
\end{figure}

Averaging over initial and summing over final spin polarizations leads us to the unpolarized squared matrix element
\begin{align}
    |\m{M}^{\gamma}_{q\bar{q}}|^2 &=\frac{4 g_s^4}{q^4} \Omega^{\gamma}_{q\bar{q}} \, \frac{C_F}{2} A^{\mu \nu}
    B_{\mu \nu} \equiv \Omega^{\gamma}_{q\bar{q}} |\m{S}^{\gamma}_{q\bar{q}}|^2\, ,
\end{align}
with the antenna current $A^{\mu \nu}$
\begin{align}
     A^{\mu \nu}(p,p',s,\bar{s})&= \frac{s^{\mu}s^{\nu}}{r^4}-2\frac{\bar{s}^{\mu}s^{\nu}}{\bar{r}^2 r^2}+\frac{\bar{s}^{\mu}\bar{s}^{\nu}}{\bar{r}^4} \, ,
\end{align}
and the recoil quark current $B^{\mu \nu}$
\begin{align}
    B^{\mu \nu}(p,p')&=\frac{1}{2}\Tr (\slashed p \gamma^{\mu} \slashed p' \gamma^{\nu})= \nonumber \\
    &=2(p^{\mu}p'^{\nu}+p^{\nu}p'^{\mu}-g^{\mu \nu}p\cdot p') \, .
\end{align}
The antenna production term $\Omega^{\gamma}_{q \bar{q}}$ depends solely on $\Gamma^{\gamma}_{q \bar{q}}$.
We now separate the antenna current $A^{\mu \nu}$ into two terms, $A^{\mu \nu}=A^{\mu \nu}_s+A^{\mu \nu}_{\bar{s}}$, with 
\begin{align}
    A^{\mu \nu}_s&\equiv\frac{s^{\mu}s^{\nu}}{r^4}-\frac{\bar{s}^{\mu}s^{\nu}}{r^2\bar{r}^2}
\end{align}
and analogously for $A^{\mu \nu}_{\bar{s}}$ by interchanging $s$ and $\bar{s}$. We are at a reference frame where the target quark $p$ is at rest, with rest mass $m$. We take $m$ to be the smallest scale in the problem. The energy of the recoiling quark $p'$ is $\sqrt{E^2+m^2}$. By taking the $m\rightarrow 0$ limit, we can approximate the propagators as
\begin{align}
    q^2&=(p'-p)^2\approx -2 E m+\m{O}(m^2) \nonumber \\
    r^2&=q^2+2q\cdot s \approx 2 p'\cdot s +\m{O}(m) \nonumber \\
    \bar{r}^2&=q^2+2q\cdot \bar{s} \approx 2 p'\cdot \bar{s} +\m{O}(m) \, .
\end{align}
Contraction of one of the antenna currents with the current $B^{\mu \nu}(p,p')$ yields
\begin{align}
    W_s &\equiv A^{\mu \nu}_s B_{\mu \nu}= \nonumber \\
    &=\frac{1}{2} \left(
    \frac{p\cdot s}{p'\cdot s}- \frac{p\cdot \bar{s}}{p'\cdot \bar{s}}+\frac{(p\cdot p')(s\cdot \bar{s})}{(p'\cdot s)(p' \cdot \bar{s})}
   \right) \, .
\end{align}
We readily recognize the well-known pattern found in the context of soft gluon emission~\cite{Mueller:1981ex,Ermolaev:1981cm}, where the recoil quark $p'$ plays the role of the emitted gluon
\begin{align}
    W_s =\frac{m}{2E} \Bigl(
    &\frac{1}{1-\cos \theta_{p's}}- \frac{1}{1-\cos \theta_{p'\bar{s}}} \nonumber \\
    & +\frac{1-\cos \theta_{s\bar{s}}}{(1-\cos \theta_{p's})(1-\cos \theta_{p'\bar{s}})}\Bigr) \, .
\end{align}
$W_s$ ($W_{\bar{s}}$) possesses a collinear divergence only when the angle between the recoil quark and the quark (anti-quark) leg goes to zero, i.e. $\theta_{p's}(\theta_{p'\bar{s}})\rightarrow 0$, and can be reinterpreted as describing the interaction with just the quark (anti-quark) leg.
By aligning the reference frame with respect to the quark $s$, and integrating the recoil quark $p'$ over the azimuthal angle $\phi$ one gets
\begin{align}
\label{eq:angord}
    \int \frac{d\phi}{2\pi}W_s=\frac{m}{E}\frac{1}{1-\cos \theta_{p's}}\Theta(\theta_{s\bar{s}}-\theta_{p's}) \, ,
\end{align}
enforcing the recoil quark angle to lie within a cone of the size of the angle of the dipole centered around the quark leg. As long as the dipole stays in a color coherent state, the successive \emph{recoils produced by soft scatterings draw the shape of the antenna}.
These results are illustrated in Fig.~\ref{fig:sketch}.
Perhaps an interesting fact is that, in contrast to the soft emission calculation, the squared of the individual matrix elements is non-vanishing and naturally provides the necessary terms to construct $W_s$ and $W_{\bar{s}}$ \footnote{Soft emission patterns are recovered if in Eq.~\eqref{eq:photmatrix} the quark current is replaced by the real emission of a gluon with momentum $q = p'-p$:
\begin{equation*}
    i g_s a_{ij} B_{\mu} \rightarrow \epsilon^*_{\mu}(q) \, .
\end{equation*}
}.

Just as in the soft emission scenario, these physics can be heuristically discussed in terms of the so-called Chudakov effect~\cite{chudakov} in cosmic ray physics. By virtue of the Heisenberg uncertainty principle, the lifetime of the virtual quark carrying momentum $r=s+q$ is $\Delta t\sim 1/\Delta E$, with $\Delta E=|\mathbf{s}|+|\mathbf{q}|-|\mathbf{r}| \sim E \theta^2=\theta/\lambda_T$, with $\lambda_T$ the wavelength of the recoil quark transverse to the dipole quark. In order to resolve the dipole internal structure, $\lambda_T$ needs to be smaller than the dipole opening $d$ after the virtual quark lifetime, so $d=\theta_{q\bar{q}}\Delta t=\theta_{q\bar{q}}\lambda_T/\theta>\lambda_T$, and then $\theta<\theta_{q\bar{q}}$.
Since in the decay of the virtual photon the total charge of the dipole is zero, recoil quarks with angles $\theta>\theta_{q\bar{q}}$ are forbidden. If the dipole is produced by the decay of a charged object, such as a quark or a gluon, then these large angle scatterings are allowed and are proportional to the corresponding Casimir. The reader will find the computations for the scattering of a soft quark with  the $g\rightarrow q \bar{q}$, $q\rightarrow qg$ and $g\rightarrow gg$ antennas in App.~\ref{sec:app}.

\section{Collisional Energy Loss of A Color Singlet Dipole}
\label{sec:collloss}

A quantity of interest that we can now compute is the collisional energy loss rate of a color dipole. As a simple test case, we do so for the dipole produced by the decay of a $Z$ boson, which closely resembles that of the virtual photon since it corresponds to a color singlet antenna. The matrix element reads
\begin{align}
&i \m{M}^{Z}  =  \frac{-2 g_s^2}{q^2}a_{ij}a_{kl}B^{\mu}\left( \frac{s_{\mu}}{r^2}-\frac{\bar{s}_{\mu}}{\bar{r}^2}\right)  \, \Gamma^{Z}_{q\bar{q}}\, ,
\end{align}
where $\Gamma^{Z}_{q\bar{q}} = ig_Z\bar{u}(s)\slashed\epsilon(k)(g_V-g_A\gamma^5)v(\bar{s})$, with the electroweak couplings
\begin{equation}
    g_Z=\frac{e}{\sin\theta_W\cos\theta_W}\,,\;\;g_V=\frac{I_3}{2}-Q\sin^2 \theta_W\,,\;\;g_A=\frac{I_3}{2}\,,
\end{equation}
where $I_3$ is the isospin of the quarks forming the antenna and $\theta_W$ is the Weinberg angle of the electroweak interaction. The unpolarized squared matrix element is
\begin{align}
    |\m{M}^{Z}_{q\bar{q}}|^2 =& \Omega^{Z}_{q\bar{q}} |\m{S}^{Z}_{q\bar{q}}|^2\, ,
\end{align}
with $|\m{S}^{Z}_{q\bar{q}}|^2=|\m{S}^{\gamma}_{q\bar{q}}|^2$ and
\begin{align}
      \Omega^{Z}_{q \bar{q}}=&g_Z^2\Tr (\slashed s \gamma^{\mu}(g_V-g_A\gamma^5) \slashed{\bar{s}}(g_V-g_A\gamma^5) \gamma^{\nu}) \nonumber\\ 
      &\times \frac{1}{3} \sum_{\lambda}\epsilon_{\mu}^{\lambda}(k)\epsilon^{\lambda *}_{\nu}(k) = \nonumber \\
    =& \frac{1}{3}g_Z^2 (g_V^2-g_A^2)\Tr(\slashed s \gamma^{\mu}\slashed{\bar{s}}\gamma^{\nu})\left( -g_{\mu\nu}+\frac{k_\mu k_\nu}{k^2} \right)\approx\nonumber\\
    \approx& \frac{8}{3}g_Z^2 (g_V^2-g_A^2)\,s\cdot \bar s \,.
\end{align}

The differential cross-section of the whole process factorizes into the decay rate of the $Z$ boson into a $q\bar{q}$ antenna, $d\Gamma_Z$, and the scattering process between the color dipole and the quark, as
\begin{align}
    d\sigma = d\Gamma_Z\frac{2w}{4F} |\m{S}^{Z}_{q\bar{q}}|^2 \frac{d^3p'}{(2 \pi)^3 2 E} \, ,
\end{align}
where $F=m\sqrt{w^2-M^2}$, with $M\approx \sqrt{2 s\cdot\bar{s}}$ the mass of the $Z$. The $Z$ decay rate is
\begin{align}
\label{eq:zdecay}
    d\Gamma_Z=\frac{\Omega_{q\bar{q}}^{Z}}{2w}\frac{d^3s}{(2 \pi)^3 2 w_s}\frac{d^3\bar{s}}{(2 \pi)^3 2 w_{\bar{s}}}(2\pi)^4\delta^{(4)}(s+\bar{s}-k) \, .
\end{align}

Taking the term that is interpreted as the interaction with the quark leg of the antenna, we write
\begin{align}
    d\sigma_q=d\Gamma_Z\frac{w}{2F}\frac{C_F}{2}\frac{4 g_s^4}{q^4} d\theta\frac{\sin \theta}{16 \pi^3} W_s d\phi E dE  \, ,
\end{align}
where $\theta$ is the angle between the recoil quark and the quark leg.
Performing the integration in $\phi$ yields
\begin{align}
    d\sigma_q&=d\Gamma_Z\frac{w}{F}\frac{C_F}{2}\alpha_s^2\frac{dE}{mE^2}d\theta\cot(\theta/2)\Theta(\theta_{q\bar{q}}-\theta) \, ,
\end{align}
where we have used the result from Eq.~\eqref{eq:angord} and $q^4 \approx 4 m^2E^2$. Note that the soft divergence associated to the propagator $q^4$, $E^{-2}$, actually corresponds to the typical angular divergence of a Rutherford scattering, since no exchanged momentum ($E\rightarrow 0$) would translate into zero deflection of the corresponding leg of the antenna.
Comparing the final soft divergence of the cross-section, $dE/E^2$, to the soft divergence of gluon emission, $dE/E$, we understand the difference as arising from an extra factor of $1/E$ from Eq.~\eqref{eq:angord}, which is absent in the soft gluon emission computation.

The scattering rate $d\Gamma_q$ is equal to the cross-section times the density of scatterers $n$, so $d\Gamma_q=n d\sigma_q$. The total (accounting for the two legs of the dipole) average energy loss per unit length $dE_T/dx$ is thus   
\begin{align}
  &\frac{dE_T}{dx}=2\langle E\,  d\Gamma_q \rangle = \nonumber \\
  &=2d\Gamma_Z\frac{w}{F}\frac{C_F}{2}\frac{n}{m}\m{A}(\theta_0,\theta_{q\bar{q}}) \m{E}(E_{\rm M},E_{\rm m})\, ,
\end{align}
where the integral over the recoil energy $E$ over a minimum energy ($E_m$, which relates to the angular cutoff of the $t$-channel propagator) and maximum energy ($E_M \ll w$) is
\begin{align}
    &\m{E}(E_{\rm M},E_{\rm m})\equiv \int_{E_{\rm m}}^{E_{\rm M}}\frac{dE}{E}\alpha_s^2(|q|)= \nonumber \\
    &=\frac{16 \pi^2}{\beta_0^2}\frac{\log \frac{E_{\rm M}}{E_{\rm m}}}{\log \frac{2E_{\rm m}^2}{Q_0^2}\log \frac{2E_{\rm M}E_{\rm m} }{Q_0^2}} \, ,
\end{align}
and we have used the expression $\alpha_s(\mu)=2\pi/\beta_0 \log( \mu/Q_0)$ for the running of the strong coupling. 
In the small angle limit, the angular integral is
\begin{align}
    \m{A}(\theta_0,\theta_{q\bar{q}})\approx 2 \log \frac{\theta_{q\bar{q}}}{\theta_0} \, .
\end{align}
This is the divergence associated to the virtual propagators in the dipole, occurring when the recoil is collinear with the corresponding leg, and is absent when one considers an elastic scattering between an on-shell parton, coming from infinity, and the soft quark~\cite{bjorkenquench}. Retaining the masses of the quarks of the antenna in the propagator naturally regulates this divergence. In the case of soft gluon emissions, this mass singularity is regulated by the inclusion of the virtual contributions. A more careful discussion of the treatment of these divergences for the case of dipole-quark scattering, possibly including the so-called contact terms that result into a zero-point subtraction prescription, is left for future work.

The restriction of the angular phase-space of the recoiling quark reduces the amount of collisional energy loss for narrower antennas as compared to wider ones. This means that during the decoherence time, $\tau_d$, 
while stimulated emissions are strongly suppressed for a color singlet dipole,
\emph{collisional energy loss is sensitive to the details of the dipole substructure}. 
As per the effects of collisional energy loss solely, one thus expects that those $Z$ bosons with decay products that are widely separated will lead, on average, to smaller values of the $Z$ reconstructed mass (ignoring medium-induced broadening effects) than those with narrower angles. These features can be analyzed by expressing the $Z$ decay rate, Eq.~\eqref{eq:zdecay}, in terms of the angle and energies of the decay products, i.e. $d\Gamma_Z/d\theta_{q\bar{q}}dw_sdw_{\bar{s}}$.
A detailed phenomenological study that can complement the proposal to measure the shift of the $W$ boson mass due to radiative energy loss~\cite{Apolinario:2017sob} is left to future work.

\section{Summary and Outlook}
\label{sec:conc}

In this Letter we have computed the cross-section for the scattering process between a color dipole and a quark at rest \footnote{One could have also considered the scattering with a gluon, which acquires a screening mass in a gauge plasma~\cite{Kapusta:1979fh}.}. We have found that, in the soft limit, i.e. when $E\ll w$, with $E$ the energy of the recoiling quark and $w$ the total energy of the dipole, the angular phase-space of the recoiling quark depends on the opening angle of the dipole itself due to quantum interference effects. In this long wavelength limit the recoil quark behaves classically and interactions can be effectively arranged as if they took place with the different color charges of the dipole separately, provided one respects the angular constraints.
This is in complete analogy to the physics found in the context of soft gluon emission off a color dipole~\cite{Mueller:1981ex,Ermolaev:1981cm}, where the recoiling quark plays the role of the soft gluon.
Using these results, we have been able to obtain the first estimate of the collisional energy loss rate of a color-coherent dipole.

Within a dense medium, such as the QGP created in heavy-ion collisions, an energetic parton frequently experiences such soft scatterings, leading to the well-studied radiative energy loss effects. It has been common practice to treat these interactions in the limit in which the medium scattering centres are static, homogeneously distributed (with some recent notable exceptions where the ensembles of scattering centres can be inhomogeneous~\cite{Barata:2022krd,Barata:2022utc,Barata:2023qds,Barata:2023zqg,Kuzmin:2023hko} and non-static~\cite{Sadofyev:2021ohn,Andres:2022ndd,Kuzmin:2023hko}), and have infinite mass, neglecting back-reaction and thus justifying the modelling of the medium as a stochastic, classical background field. This simplification facilitates the resummation of an infinite number of these scatterings, allowing for a convenient reformulation of the problem in terms of Wilson lines. Under this framework, it has been shown that these multiple soft scatterings can lead to color decoherence between the two legs of a dipole after a timescale $\tau_d$~\cite{Mehtar-Tani:2011hma,Mehtar-Tani:2011vlz,Casalderrey-Solana:2011ule,Mehtar-Tani:2012mfa}. 
In the present work we have focused on the properties of the recoiling partons that participate in soft scatterings, which obviously means that we have had to consider back-reaction on the medium constituents. These differences in the medium modelling assumptions between all previous work, addressing dipole color decoherence, and the present work, addressing imprints of color coherence in the recoil properties, motivates further work to establish a comprehensive framework to simultaneously address both phenomena.

The clear message from the results here presented is that as long as a dipole is in a color coherent state, recoils produced in soft scatterings are restricted to lie at the angular regions allowed by color coherence effects.
Just like in the usage in vacuum parton showers of the AO phenomena of soft gluon emissions, this picture suggests a way forward towards implementing color coherence effects in the description of elastic collisions in jet quenching models. 
One could in principle consider elastic scatterings off each of the legs of a given dipole independently, by sampling a collisional rate $d\Gamma_q$, ensuring that they respect the angular constraints illustrated in, e.g., Eq.~\eqref{eq:angord}, or those presented in App.~\ref{sec:app} for the rest of possible antennas, depending on the specific color flow topology -- as long as they are in a color coherent state (namely if time $\tau<\tau_d$). A dedicated phenomenological study using jet quenching Monte Carlos will be done in future work.

The results presented in this Letter represent an important improvement in the theoretical description of high-energy jets traversing deconfined QCD media, such as the QGP created in heavy-ion collisions. A better understanding of the recoils involved in elastic scatterings with the color-coherent multi-partonic structures generated by jet evolution is crucial for characterizing medium response physics and, by extension, for understanding the nature of the QGP itself.
\\

\begin{acknowledgements}
   We are thankful to Carlota Andr\'es, N\'estor Armesto, Jo\~{a}o Barata, Jorge Casalderrey-Solana, Paul Caucal, Fabio Dom\'{i}nguez, Xabier Feal, Xo\'{a}n Mayo L\'{o}pez, Mateusz Ploskon, Krishna Rajagopal, Diego Rodr\'{i}guez-L\'{o}pez, Andrey Sadofyev, Carlos Salgado, Alba Soto-Ontoso, Adam Takacs and Konrad Tywoniuk for helpful discussions. DP is funded by the European Union's Horizon 2020 research and innovation program under the Marie Sk\l odowska-Curie grant agreement No 101155036 (AntScat), by the European Research Council project ERC-2018-ADG-835105 YoctoLHC, by Spanish Research State Agency under project PID2020-119632GB- I00, by Xunta de Galicia (CIGUS Network of Research Centres) and the European Union, and by Unidad de Excelencia Mar\'ia de Maetzu under project CEX2023-001318-M.
\end{acknowledgements}

%


\appendix

\section{$g\rightarrow q\bar{q}$, $q\rightarrow q g$ and $g\rightarrow g g$ Antennas}
\label{sec:app}

In this Section we present the computation of the matrix elements between a dipole and a soft quark for the rest of possible color configurations. 
\\

\emph{$g\rightarrow q \bar{q}$}. We start by replacing the incoming virtual photon with a virtual gluon. The two leading diagrams from the previous calculation become
\begin{align}
    i\m{M}_1^g+i\m{M}_2^g=&\frac{-2 g_s^2}{q^2}B^{\mu}a_{ij}
    \Bigl\{ 
    (a b)_{kl}\frac{s^{\mu}}{r^2}-
    (b a)_{kl}\frac{\bar{s}^{\mu}}{\bar{r}^2}
    \Bigr\} \Gamma^g_{q\bar{q}} \, ,
\end{align}
where now $\Gamma^{g}_{q\bar{q}}\equiv i g_s \,  \bar{u}(s)\Gamma v(\bar{s})$.
The new diagram we need to consider, namely the interaction with the incoming gluon, is notably simplified in the limit in which one of the legs carries soft momenta ($q \ll s+\bar{s}$), enhancing the term that preserves incoming gluon polarization. Within this approximation, it reads
\begin{align}
    i \m{M}_3^{g}&=\frac{2 g_s^2}{q^2} a_{ij}\left[a,b\right]_{kl} B^{\mu}\left( \frac{s_{\mu}+\bar{s}_{\mu}}{k^2}\right) \Gamma^g_{q\bar{q}} \, .
\end{align}
However, it is not enhanced in the soft limit, since $k^2=(s+\bar{s}+q)^2\sim \m{O}(\beta^0)$, and is thus a subleading contribution that we drop. Therefore, $i\m{M}^g_{q\bar{q}}\approx i \m{M}_1^{g}+i \m{M}_2^{g}$ to leading order in $\beta$.

Averaging over initial color states ($N$ possible colors for the incoming quark and $N^2-1$ for the incoming gluon) and summing over final color states, we recognize two types of color flows. The direct color flow terms yield
\begin{align}
\label{eq:dircol}
&\frac{1}{N(N^2-1)}\sum (a_{ij}(ab)_{kl})^* \, (c_{ij}(cb)_{kl}) =\frac{C_F}{4 N} \, .
\end{align}
The cross color flow terms yield
\begin{align}
\label{eq:crosscol}
    &\frac{1}{N(N^2-1)}\sum (a_{ij}(ab)_{kl})^* \, (c_{ij}(bc)_{kl})=\frac{C_F}{4 N}-\frac{1}{8} \, .
\end{align}

The unpolarized squared matrix element is then
\begin{align}
    |\m{M}^g_{q\bar{q}}|^2 &=\frac{4g_s^4}{q^4} \Omega^g_{q\bar{q}} \frac{1}{4N}B_{\mu \nu}\left[ C_F
   A^{\mu \nu}
    +C_A
  \frac{s^{\mu}\bar{s}^{\nu}}{r^2\bar{r}^2}
   \right]
    \, ,
\end{align}
where $\Omega^g_{q \bar{q}}$ is again associated to dipole production.
The term proportional to $C_F$ presents, as before, a recoil quark with an angle contained \emph{within} the cone of the $q\bar{q}$ antenna centered around each of the legs. The new term $s^{\mu}\bar{s}^{\nu}/r^2\bar{r}^2\equiv J^{\mu \nu}$, proportional to $C_A$, enforces a recoil angle $\theta$
\begin{align}
\label{eq:noord}
    \int \frac{d\phi}{2\pi} J^{\mu \nu}B_{\mu \nu}=\frac{m}{E}\frac{1}{1-\cos \theta}\Theta(\theta-\theta_{s\bar{s}}) \, ,
\end{align}
exclusively \emph{outside} of the cone of the dipole centered around each of the legs, as if the scattering took place with the total charge of the dipole.
\\

\emph{$q\rightarrow q g$.} The sum of the leading matrix elements yields (with $\bar{s}$ the momentum of the outgoing gluon)
\begin{align}
   & i\m{M}^{q}_{qg}=\frac{-2 g_s^2}{q^2}  B^{\mu}a_{ij}\Bigl\{ 
    (a b)_{kl}\frac{s_{\mu}}{r^2}-[a,b]_{kl}\frac{\bar{s}_{\mu}}{\bar{r}^2}
    \Bigr\}\Gamma^q_{qg} \, ,
\end{align}
where we have defined $\Gamma^q_{qg}\equiv i g_s \bar{u}(s) \epsilon^*_{\alpha}(\bar{s})\Gamma^{\alpha}$. Again, the diagram that describes the interaction with the incoming virtual quark is subleading in $\beta$ and is neglected.

The unpolarized squared matrix element is
\begin{align}
    |\m{M}^{q}_{qg}|^2 &=\frac{4 g_s^4}{q^4} \Omega^q_{qg} B_{\mu \nu}\nonumber \frac{C_F}{2N} \left[ C_F
    \frac{s^{\mu}s^{\nu}}{r^4}
    +C_A A_{\bar{s}}^{\mu \nu}
   \right]
    \, ,
\end{align}
with $\Omega^q_{qg}$ associated to the production of the dipole.
We observe that there is no angular restriction for the $C_F$ term (as if the scattering took place with the quark leg of the dipole, which is also the total charge of the dipole), while in the $C_A$ term coherence restricts the angular distribution of the recoil to be within the cone of the dipole centered around the gluon (as if the scattering took place with the gluon leg of the dipole). 
\\

$g\rightarrow g g$.
In the soft limit, the sum of the two leading matrix elements is simply (with $s^{\mu}$ and $\bar{s}^{\mu}$ the momenta of the outgoing gluons)

\begin{align}
   & i\m{M}^{g}_{gg}=\frac{2i g_s^2}{q^2} \Gamma^g_{gg} B_{\mu}a_{ij}\Bigl\{ 
    \frac{s^{\mu}}{r^2}f^{abc}f^{bed}
    +\frac{\bar{s}^{\mu}}{\bar{r}^2}f^{abe}f^{bdc}
    \Bigr\} \, ,
\end{align}
with $\Gamma^g_{gg}\equiv g_s \epsilon^*_{\alpha}(s)\epsilon^*_{\beta}(\bar{s})\Gamma^{\alpha \beta}$. The direct color flow terms yield
\begin{align}
&\frac{1}{N(N^2-1)}\sum (a_{ij}f^{abc}f^{bed})^* \, (g_{ij}f^{ghc}f^{hed})=\frac{C_A}{2} \, ,
\end{align}
and the cross ones
\begin{align}
&\frac{1}{N(N^2-1)}\sum (a_{ij}f^{abc}f^{bed})^* \, (g_{ij}f^{ghe}f^{hdc})=-\frac{C_A}{4} \, .
\end{align}

The unpolarized squared matrix element is then
\begin{align}
    |\m{M}^{g}_{gg}|^2 &=\frac{4 g_s^4}{q^4}\Omega^g_{gg} B_{\mu \nu} \frac{C_A}{2}\left[
    \frac{s^{\mu}s^{\nu}}{r^4}+
\frac{\bar{s}^{\mu}\bar{s}^{\nu}}{\bar{r}^4}-\frac{s^{\mu}\bar{s}^{\nu}}{r^2\bar{r}^2} 
   \right] =
    \nonumber \\
    &=\frac{4 g_s^4}{q^4} \Omega^g_{gg} B_{\mu \nu}\frac{C_A}{2}
    \left[
    A^{\mu \nu}+J^{\mu \nu}
    \right] \, ,
\end{align}
where $A^{\mu \nu}$ enforces scattering angles within a cone centered around each of the gluon legs and $J^{\mu \nu}$ corresponds to out-of-cone recoil angles, as expected from the total charge of the antenna.

We can compactly express all previous results in the following way~\cite{Mehtar-Tani:2011lic}
\begin{align}
    |\m{M}_i|^2&=\frac{4 g_s^4}{q^4} \Omega_i \, C_i \, B_{\mu \nu}  \nonumber \\
   &\times  \left(
    u^{\mu}u^{\nu} Q_u^2+v^{\mu}v^{\nu} Q_v^2+2 u^{\mu}v^{\nu}Q_u\cdot Q_v
    \right) \, ,
\end{align}
where $u$ and $v$ represent the two legs of the antenna $u^{\mu}\equiv s^{\mu}/r^2$, $v^{\mu}\equiv \bar{s}^{\mu}/\bar{r}^2$. The corresponding charge vectors satisfy $Q_u+Q_v=Q_t$, where $Q_t$ is the charge of the incoming virtual object, $Q_q^2=Q_{\bar{q}}^2=C_F$ and $Q_g^2=C_A$. By solving for $Q_u\cdot Q_v$ using the equation $(Q_u+Q_v)^2=Q_t^2$ one recovers all the possible color configurations. The index $i$ labels the four different cases, for which we have different dipole production terms $\Omega_i$ and different global color factors $C_i=\lbrace 1/2,1/4N,C_F/2N,1/2\rbrace$ for processes $i=\lbrace \gamma q\bar{q},gq\bar{q},qqg,ggg \rbrace$, respectively.

\end{document}